\documentclass[apj,numberedappendix]{emulateapj}
\usepackage{apjfonts}

\slugcomment{Submitted for publication in The Astrophysical Journal Letters}
\shorttitle{SHEAR WAVES AND GIANT FLARE OSCILLATIONS}
\shortauthors{PIRO}

\newcommand{\be}{\begin{eqnarray}}
\newcommand{\ee}{\end{eqnarray}}

\begin{document}


\title{Shear Waves and Giant Flare Oscillations from Soft Gamma-Ray Repeaters}

\author{Anthony L. Piro} 
\affil{Department of Physics, Broida Hall, University of California, Santa Barbara, CA 93106; piro@physics.ucsb.edu}


\begin{abstract}

  Recent observations of giant flares from soft gamma-ray repeaters
have exhibited multiple $\sim25-150\ {\rm Hz}$ oscillations. Frequencies in this
range are expected for
toroidal shear waves in a neutron star (NS) crust, lending support to Duncan's proposal
that such modes may be excited in these events. This motivates a reassessment of how
these waves reflect the NS structure and what role the magnetic field plays in setting
their frequencies. We calculate the eigenfrequencies and eigenfunctions of
toroidal oscillations for a realistic NS crust, including a vertical magnetic field
at magnetar strengths ($B\sim10^{14}-10^{15}\ {\rm G}$). The lowest radial-order
mode has a red-shifted frequency of $\approx28\ {\rm Hz}[l(l+1)/6]^{1/2}$, with
the prefactor depending on the NS's mass and radius, and its crust's depth and composition.
This mode is independent of the magnetic field for $B\lesssim4\times10^{15}\ {\rm G}$, a limit
much greater than the inferred dipole magnetic fields for these objects.
Though this is a good fit to the observed oscillations, only rather loose
constraints can be made for the NSs' properties because all that can be fit is this
prefactor (a single parameter). Modes with shorter radial wavelengths are
more sensitive to the magnetic field starting at $B\sim2\times10^{14}\ {\rm G}$
and have higher frequencies ($\sim600-2000\ {\rm Hz}$).
The discovery of these modes, coupled
with the oscillations observed thus far, would
provide a powerful probe to the NS crustal structure.
\end{abstract}



\section{Introduction}

  Soft gamma-ray repeaters (SGRs) are identified by their
short ($\sim0.1\ {\rm s}$), recurrent bursts of soft gamma-rays
\citep[see review by][]{wt04}.
Typical bursts reach luminosities of $\sim10^{41}\ {\rm ergs\ s^{-1}}$,
considerably above the Eddington limit. All four of the known
SGRs have spin periods of $\approx5-8\ {\rm s}$, and three have
large period derivatives (with respect to typical radio pulsars) of
$\sim10^{-10}\ {\rm s\ s^{-1}}$, implying dipole magnetic fields of
$B\approx(5-8)\times10^{14}\ {\rm G}$ \citep{wt04}. SGRs have
been successfully
explained by the ``magnetar'' model \citep{dt92,td95} in which these
objects are isolated neutron stars (NSs) powered by the decay of ultra-strong
magnetic fields.

  In addition to the shorter bursts, there have been three ``giant
flares'' that released $\sim10^{44}-10^{46}\ {\rm ergs}$ of energy,
which took place 1979 March 5 from SGR $0526-66$
\citep{maz79}, 1998 August 27 from SGR $1900+14$
\citep{hur99,fer99},
and 2004 December 27 from SGR $1806-20$
\citep{hur05,pal05}. In addition to the large modulations
seen at each object's known spin, all
three of the giant flares showed signs of periodicities at much higher
frequencies. From SGR $1806-20$, \citet{isr05} reported
a strong oscillation at $92.5\ {\rm Hz}$ with weaker oscillations at $18$
and $30\ {\rm Hz}$. \citet{sw05} discovered a $84\ {\rm Hz}$ oscillation
in SGR $1900+14$ with weaker oscillations at $53.5$ and $155.1\ {\rm Hz}$
and a possible additional feature at $28\ {\rm Hz}$. Finally, there
was some hint of a $\approx43.5\ {\rm Hz}$ signal in the giant flare
from SGR $0526-66$ \citep{bar83}. The thrilling implication is that
these oscillations may be shear modes, excited from crustal deformations
during the flare \citep{dun98}.
If true, they would be the first modes ever seen from a NS, initiating
a new era where seismology is used to learn about NS crusts.

  By extrapolating the calculations of \citet{mvh88} with better estimates for
a NS mass, radius, and shear modulus,
\citet{dun98} provided a frequency estimate for toroidal shear modes
of a magnetar. Magnetic effects were included by assuming that
it acts as a tension
adding to the shear modulus isotropically.
\citet{sw05} showed that this estimate matches SGR $1900+14$'s
frequencies for $l=2, 4, 7,$ and $13$ (where $l$ is
the angular quantum number). In this present work we
calculate the modes'
dependence on the properties of the NS and its crust, and we provide an analytic
frequency estimate (eq. [\ref{eq:omega}]). We
investigate how and when a vertical magnetic field plays an important role in determining
the toroidal frequencies.

  In \S \ref{sec:modes} we present equations for toroidal oscillations
in a plane-parallel layer, and in \S \ref{sec:analytic} we
estimate the expected mode properties.
Numerical calculations of the modes using a realistic crust model are in \S 4.
We conclude in \S \ref{sec:conclusions}
with a summary our work and a discussion of future studies.


\section{Equations for Toroidal Shear Waves}
\label{sec:modes}

  We assume a NS with mass $M=1.4M_\odot$ and radius $R=12\ {\rm km}$.
The top of the NS's crust is at the depth where $\Gamma\equiv Z^2e^2/ak_{\rm B}T\approx173$
\citep{fh93}, where $Z$ is the charge per ion and $a=(3/4\pi n_i)^{1/3}$ is the average
ion spacing, with $n_i$ the ion number density. This typically occurs at a density
\be
	\rho_{\rm top} = 2.3\times10^9 {\rm g\ cm^{-3}}\left(\frac{T_8}{3}\right)^3\left(\frac{26}{Z}\right)^6
				\left( \frac{A}{56}\right),
\ee
where $T_8\equiv T/10^8\ {\rm K}$, $A$ is the number of nucleons per ion, and
we use Fe as the fiducial nucleus. This depth is similar to that found
for accreting NSs,
rather than like isolated ones ($\rho_{\rm top}\sim10^5\ {\rm g\ cm^{-3}}$), because
magnetars are young ($\sim10^3\ {\rm yrs}$)
and hot from magnetic field decay and recurrent flares \citep{td96}.
At $\rho\approx4\times10^{11}\ {\rm g\ cm^{-3}}$ neutron drip begins, at which
point $A$ increases rapidly with depth and neutrons play an increasingly
important role in providing pressure. The crust/core interface is at
$\rho\approx10^{14}\ {\rm g\ cm^{-3}}$ with $A\sim300-400$ \citep{dh01}, below
which the NS is liquid once again. For this reason, the crust can be thought of
as a cavity where shear oscillations can preside. 
  Since the pressure scale height at this depth,
$H = P/\rho g \approx 3\times10^4\ {\rm cm}$, is much less than the radius,
we use a plane-parallel geometry with constant gravitational acceleration,
$g=GM/R^2\approx1.3\times10^{14}\ {\rm cm\ s^{-2}}$. We use
a Cartesian coordinate system with $z$ as the radial coordinate.

  We focus on shear waves, which
are classified as toroidal oscillations \citep{as77}.
These are incompressible with no vertical
displacement, so we can write a Lagrangian displacement as
$\mbox{\boldmath$\xi$}=\xi_x\mbox{\boldmath$\hat{x}$}+\xi_y\mbox{\boldmath$\hat{y}$}$,
with $\mbox{\boldmath$\nabla\cdot\xi$}=0$.
We assume that the magnetic field is constant and oriented in the
vertical direction ($\mbox{\boldmath$B$}=B\mbox{\boldmath$\hat{z}$}$). Though this is
a drastic simplification compared to the tangled fields that may exist within a magnetar's crust,
this provides a simple first step for combining shear and magnetic effects.

  Within the crust, the shear modulus is \citep{stretal91},
\be
	\mu = \frac{0.1194}{1+0.595(173/\Gamma)^2}
		\frac{n_i(Ze)^2}{a}.
\ee
This provides a shear stress tensor \citep{ll70}, which for
toroidal displacements simplifies to be,
\be
	\delta\sigma_{ij} = \mu\left( \frac{\partial\xi_i}{\partial x_j}+ \frac{\partial\xi_j}{\partial x_i}\right).
\ee
Assuming an oscillatory time
dependence, $\xi\propto e^{i\omega t}$, where $\omega$ is the mode
frequency in a frame on the NS surface, momentum conservation in the $x$- and
$y$-directions is,
\be
	-\rho \omega^2 \xi_i = \frac{\partial}{\partial x_j}\delta\sigma_{ij}
		+\frac{1}{c}\left( \mbox{\boldmath$\delta j\times B$}\right)_i,
	\label{eq:momentum}
\ee
where we have used the Einstein summation convention for repeated indices.
The perturbed current is
\be
	\mbox{\boldmath$\delta j$}
		= \frac{c}{4\pi}\left[ \mbox{\boldmath$\nabla\times\delta B$}
			- \frac{\omega^2}{c^2}\mbox{\boldmath$\xi\times B$} \right],
	\label{eq:current}
\ee
where
$\mbox{\boldmath$\delta B$} =\mbox{\boldmath$\nabla\times$}\left(\mbox{\boldmath$\xi\times B$}\right)$ is the perturbed magnetic field.
Combining the above equations, we derive the wave equation for magnetic
toroidal shear modes,
\be
	-\rho\omega^2\mbox{\boldmath$\xi$} &=& \mu \nabla_\perp^2\mbox{\boldmath$\xi$} 
		+\frac{\partial}{\partial z}\left( \mu\frac{\partial\mbox{\boldmath$\xi$} }{\partial z}\right)
		+\frac{B^2}{4\pi}\left[ \frac{\partial^2\mbox{\boldmath$\xi$} }{\partial z^2}
			+ \frac{\omega^2}{c^2}\mbox{\boldmath$\xi$} \right],
	\label{eq:waveequation}
\ee
where $\nabla_\perp^2=\partial^2/\partial x^2+\partial^2/\partial y^2$, and
$\nabla_\perp^2\mbox{\boldmath$\xi$} = -l(l+1)/R^2 \mbox{\boldmath$\xi$} $ in this plane
parallel-geometry.

\section{Analytic Estimates}
\label{sec:analytic}

  We now make some analytic estimates for the expected mode properties.
A dispersion relation can be derived from equation
(\ref{eq:waveequation}) by taking
$\partial/\partial z\rightarrow ik_z$ and $\nabla_\perp\rightarrow ik_\perp$ to be the vertical and
transverse wavenumbers, respectively. Assuming $\mu\approx{\rm constant}$, this results in
\be
	\omega^2 = v_s^2\left( k_z^2+k_\perp^2\right)
		+ v_{\rm A}^2\left(k_z^2+\frac{\omega^2}{c^2}\right),
	\label{eq:dispersion}
\ee
where $v_s=(\mu/\rho)^{1/2}$ and $v_{\rm A}=B/(4\pi\rho)^{1/2}$ are the
shear and Alfv\'{e}n speeds, respectively. For low radial-order modes
$k_z^2\gg \omega^2/c^2$, so we drop the last term for our estimates.

  Equation (\ref{eq:dispersion}) has two limits, both of which are also
found numerically. The first is when $k_\perp\gg k_z$, as expected
for the lowest radial-order mode (which has no nodes in the radial direction
so that $k_z\approx0$). Its resulting dispersion relation is
$\omega^2=v_s^2k_\perp^2$. We evaluate this frequency
at the crust/core interface, using a fiducial composition
for isolated NSs \citep{dh01}. The shear
modulus at this depth is
\be
	\mu_{\rm bot} = 1.2\times10^{30}\ {\rm ergs\ cm^{-3}}\rho_{14}^{4/3}\left(\frac{Z}{38}\right)^2
	\nonumber \\
	\times\left(\frac{302}{A}\right)^{4/3}\left( \frac{1-X_n}{0.25}\right)^{4/3},
	\label{eq:mubot}
\ee
where $\rho_{14}\equiv \rho/10^{14}\ {\rm g\ cm^{-3}}$ and $X_n$ is the fraction
of neutrons. Setting $k_\perp^2=l(l+1)/R^2$ and including gravitational red-shifting
from the NS surface, the observed frequency is
\be
	\left.\frac{\omega_{\rm obs}}{2\pi}\right|_{n=0} &=& 28.8\ {\rm Hz}\ \rho_{14}^{1/6}
		\left( \frac{Z}{38} \right)\left( \frac{302}{A}\right)^{2/3}
		\nonumber
		\\
		&&\times\left( \frac{1-X_n}{0.25} \right)^{2/3}
		\left[ \frac{l(l+1)}{6}\right]^{1/2}R_{12}^{-1}
		\nonumber
		\\
		&&\times\left( 1.53 - 0.53 \frac{M_{1.4}}{R_{12}}\right)^{1/2},
	\label{eq:omega}
\ee
where $M_{1.4}\equiv M/1.4M_\odot$ and $R_{12}\equiv R/12\ {\rm km}$, and $n$ denotes
the number of radial nodes for this mode. This frequency
estimate is within $\approx5\%$ of what we find numerically. Notice that it
is strongly dependent on the crustal composition, but weakly
dependent on the density.

  The fact that this mode is independent of the magnetic field is clearly a result of
our simple field geometry, so it would be useful to have an estimate for the effects of
a field at an angle $\alpha$ from vertical. This would interact with
the transverse wavenumber as the shear modulus does, giving
$\omega^2\approx \left( v_s^2 + v_{\rm A}^2\sin^2\alpha \right) k_\perp^2$.
Using equation (\ref{eq:mubot}), the critical magnetic field such that
$v_s=v_{\rm A}\sin\alpha$ is
\be
	B_{\rm crit} = 3.8\times10^{15}\ {\rm G}\frac{\rho_{14}^{2/3}}{\sin\alpha}
		\left( \frac{Z}{38}\right)
		\left( \frac{302}{A}\right)^{2/3}
		\left(\frac{1-X_n}{0.25} \right)^{2/3}
	\label{eq:bcrit}
\ee
When $B>B_{\rm crit}$, we cannot trust our frequency estimate in equation
(\ref{eq:omega}).

  The other important limit is when $k_z\gg k_\perp$ as is expected for all higher radial-order
modes ($n>0$) since they have $k_z\sim1/H$. As long as $l\lesssim R/H$, this
results in
\be
	\omega^2=(v_s^2+v_A^2)k_z^2,
	\label{eq:otherfreq}
\ee
which shows that such modes are independent of $l$ and
sensitive to $B$. The vertical wavenumbers of these modes are
given by the WKB limit, i.e. $\int k_z dz\approx n\pi$ where the integral is across the
crust. This means that $B$ can have an important effect when $v_{\rm A}>v_s$
in the outer crust, even if $v_{\rm A}\ll v_s$ at the crust base.
It also shows that there should
be a frequency ratio between the $n=0$ and
$n>0$ modes on the order of $k_x/k_\perp\sim R/H\sim20-30$
\citep[as confirmed by our numerical calculations, also see][]{car86,mvh88}.
The observation and measurement of this ratio could therefore be used to constrain
the depth of the crust.

\section{Numerical Calculations}
\label{sec:calculations}

\begin{figure}
\epsscale{1.15}
\plotone{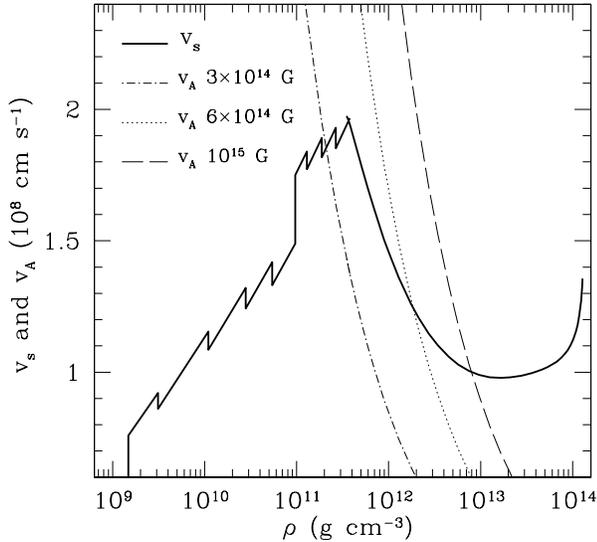}
\caption{The shear speed, $v_s$ ({\it solid line}), and the Alfv\'{e}n speed, $v_{\rm A}$,
using $B=3\times10^{14}$ ({\it dot-dashed line}),
$\ 6\times10^{14}$ ({\it dotted line}), and $10^{15}\ {\rm G}$ ({\it dashed line}).
We use the crustal compositions
of \citet{hp94} for $\rho<4\times10^{11}\ {\rm g\ cm^{-3}}$ and
\citet{dh01} for $\rho>4\times10^{11}\ {\rm g\ cm^{-3}}$ .}
\label{fig:speeds}
\epsscale{1.0}
\end{figure}

  We next perform numerical calculations using the crustal
models of \citet{hp94} above neutron drip ($\rho\approx4\times10^{11}\ {\rm g\ cm^{-3}}$)
and \citet{dh01} for $\rho>4\times10^{11}\ {\rm g\ cm^{-3}}$.
Since observed magnetars are young ($\sim10^3\ {\rm yrs}$)
and hot these cold,
beta-equilibrium models may not be entirely accurate. An important future calculation
would be to understand the magnetar crustal structure as a function of age.
The most relevant quantities for our calculations are $v_s$ and $v_{\rm A}$,
which we plot in Figure \ref{fig:speeds}.
The key point is that at the base of the crust $v_s\gg v_{\rm A}$, even
for a magnetic field considerably larger than the dipole fields of
$B\approx(5-8)\times10^{14}\ {\rm G}$ inferred for these SGRs \citep{wt04}.

  We set $\xi/R=1$ and the horizontal ``traction'' to zero in the ocean
(at $\rho\approx5\times10^7\ {\rm g\ cm^{-3}}$)
and integrate equation (\ref{eq:waveequation})
deeper, shooting for zero horizontal traction at the crust
base to find $\omega$
\citep[the traction includes both shear and magnetic contributions, for a complete
discussion see][]{car86}.
The top boundary condition is equivalent to setting
$\mbox{\boldmath$\delta B$}=0$, and its placement
was found to not change our conclusions.
The bottom boundary condition implies that the crust is decoupled
from the core, a poor approximation if the
magnetic field is as strong as $B_{\rm crit}$. Future studies should
investigate what limit this places.
We make sure that the horizontal traction is continuous at the ocean/crust
and neutron drip boundaries.

\begin{figure}
\epsscale{1.15}
\plotone{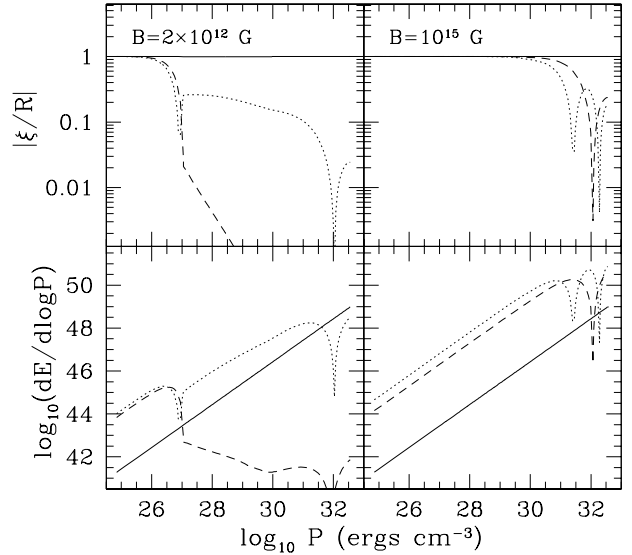}
\caption{The eigenfunctions and energy density per logarithm pressure
(measured in ergs) of the
three lowest order
modes with $l=2$. The left panel displays
$B=2\times10^{12}\ {\rm G}$ and modes with red-shifted frequencies of 27.4 ($n=0$, {\it solid line}),
521.5 ($n=1$, {\it dashed line}), and 637.2 Hz ($n=2$, {\it dotted line}). The right panel displays
$B=10^{15}\ {\rm G}$ and modes with observed frequencies of 27.4 ($n=0$, {\it solid line}),
742.8 ($n=1$, {\it dashed line}), and 1329.2 Hz ($n=2$, {\it dotted line}).}
\label{fig:modes}
\epsscale{1.0}
\end{figure}
  Figure \ref{fig:modes} shows the eigenfunctions
for the three lowest order modes with
$B=2\times10^{12}\ {\rm G}$ ({\it left panels}) and $10^{15}\ {\rm G}$
({\it right panels}), all using $l=2$.
The absolute value of the displacements are plotted, so cusps correspond
to nodes. The $n=0$ mode has no nodes and a red-shifted frequency of $27.4\ {\rm Hz}$,
very
close to equation (\ref{eq:omega}).
The $n>0$ modes are combination Alfv\'{e}n/shear waves.
When $B<(4\pi\mu)^{1/2}$ at the top of the crust, the crust acts as a hard boundary,
trapping Alfv\'{e}n waves in the ocean. As $B$ increases, this boundary becomes irrelevant and
the modes go all the way to the crust base. Thus the $n=1$ trapped ocean Alfv\'{e}n wave
at $B=2\times10^{12}\ {\rm G}$ becomes a shear wave at $B=10^{15}\ {\rm G}$.
We also plot the energy density per logarithm pressure, which
provides intuition of the depths for which the mode is most sensitive,
\be
	\frac{dE}{d\log P} = 2\pi R^2l(l+1)\omega^2\xi^2P/g.
\ee
The $n=0$ mode's energy is concentrated at the crust base,
so it is most sensitive to this depth, explaining
why our analytic estimate is so accurate.
The $n>0$ modes have $dE/d\log P$ more constant across the crust,
so they are sensitive to when
$v_{\rm A}\gtrsim v_s$ in the outer regions.

  The energy density allows us to test whether the excitation of these modes is
plausible. A giant flare releases $\sim10^{44}-10^{46}\ {\rm ergs}$ of energy,
while from Figure \ref{fig:modes} we estimate that $\sim10^{49}\ {\rm ergs}$
is needed to excite the $n=0$ mode to $\xi\sim R$. Since the mode energy scales
$\propto \xi^2$, $10^{43}\ {\rm ergs}$ is enough energy for
$\xi\sim0.001R\sim10^3\ {\rm cm}$, which is of order the pressure
scale height at the top of the crust. It seems likely that a small amount of the
energy from a flare can excite appreciable amplitudes. In contrast, the $n>0$
modes require orders of magnitude more energy for a similar size, perhaps
explaining why they have yet to be detected.

\begin{figure}
\epsscale{1.15}
\plotone{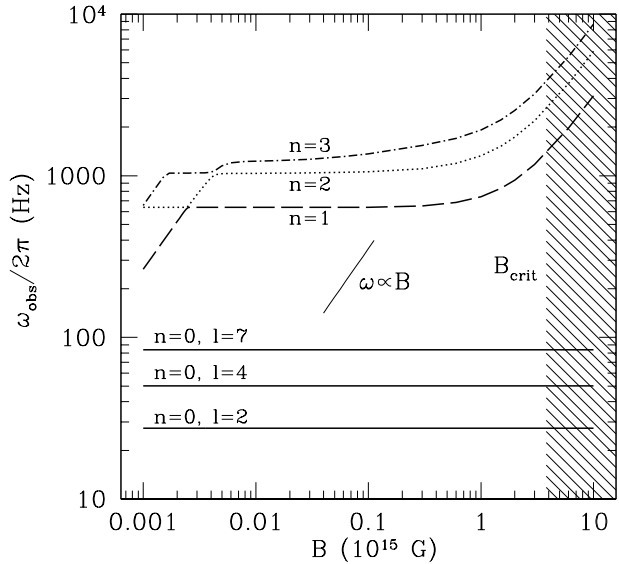}
\caption{The observed frequencies of the $n=0$ ({\it solid lines}),
1 ({\it dashed line}), 2 ({\it dotted line}), and 3 ({\it dot-dashed line}) modes
as a function of $B$. We plot $l=2, 4$, and 7 for $n=0$. The $n>0$
modes are independent of $l$ (at least at these low $l$-values).
Frequencies with $B>B_{\rm crit}$ ({\it shaded region})
cannot be trusted if the magnetic field has a non-vertical component.
Notice that at low $B$ avoided crossings occur between the trapped
ocean Alfv\'{e}n waves and shear waves.}
\label{fig:frequencies}
\epsscale{1.0}
\end{figure}

  In Figure \ref{fig:frequencies} we plot the red-shifted frequencies of the $n=0, 1, 2,$ and $3$
modes as a function of $B$. This confirms that the $n=0$ mode is not affected by the magnetic
field and scales $\propto[l(l+1)]^{1/2}$, while
the $n>0$ modes are independent of $l$ and exhibit $\omega\propto B$ when they are
Alfv\'{e}n-like in character. This happens at low $B$ for the trapped ocean
Alfv\'{e}n waves (as described above) and when $B\gtrsim2\times10^{14}\ {\rm G}$ .
The frequencies of the trapped oscillation are likely sensitive
to our top boundary condition, and not robust (whereas the
frequencies for $B\gtrsim7\times10^{12}\ {\rm G}$
are accurate because these modes are concentrated within the crust).

  The numerically calculated $n=0$, $l=2,4,7,$ and $13$ modes
have frequencies of $27.4, 50.1, 83.8,$ and $151.0\ {\rm Hz}$, respectively,
between $\approx0.2-6\%$ of SGR $1900+14$'s frequencies.
This confirms Strohmayer \& Watts' identification, and thus is strong evidence
for a shear wave interpretation of the giant flare oscillations. This does not
provide strong constraints on physical properties because the observed frequencies
scale $\propto[l(l+1]]^{1/2}$. Therefore, only the prefactor can be fit, providing a
single measured parameter, and as equation (\ref{eq:omega}) shows,
this parameter is degenerate with respect to multiple properties of the NS. The
$n>0$ modes would be useful for getting better constraints on the crustal properties,
and Figure \ref{fig:frequencies} shows that searches should be conducted in the
range of $\sim600-2000\ {\rm Hz}$. Due to the energetics arguments described above,
the powerful SGR $1806-20$ giant flare is most promising for such a discovery.
The $30$ and $92.5\ {\rm Hz}$ oscillations from SGR $1806-20$ naturally
map to $l=2$ and 7, respectively, but its $18\ {\rm Hz}$ oscillation has yet to be
identified. Though a crustal interface wave naturally provides a frequency in this range
\citep{pb05}, magnetic effects cannot be ignored at the top of the crust, so further
investigation is required.

\section{Conclusions and Discussion}
\label{sec:conclusions}

  We have calculated the properties of shear
modes with a vertical magnetic field and compared them
to the giant flare oscillations seen from SGRs. The $n=0$
mode matches the observed frequencies and is
largely independent of the magnetic field strength. This is
not surprising given that the two dominant oscillations from SGR $1900+14$
and SGR $1806-20$ are only 10\% different
($84$ and $92.5\ {\rm Hz}$, respectively), while the inferred
magnetic fields from these systems differ by a factor of $1.4$. Some tuning
of the interior field strength is required if magnetic effects
are the predominant factor setting the observed frequencies. It seems
more likely that differences in these NSs' masses, radii, or crusts (as highlighted in
eq. [\ref{eq:omega}]) are the primary reason for the different frequencies
of these two systems. Nevertheless, future studies should consider more
realistic field geometries since they can act on
the modes in surprising ways \citep{mps01}.

  Even if the magnetic field does
not play a primary role in setting the $n=0$ frequencies, it is likely
crucial in allowing the oscillations to be observed. Toroidal oscillations are incompressible
so they are normally decoupled from the ocean and do not produce temperature
and intensity variations. The magnetic field changes things;
it creates trapped ocean Alfv\'{e}n modes and causes the shear waves
to have considerable amplitudes throughout the ocean. A calculation
of the transmission through the ocean and magnetosphere \citep[e.g.,][]{bla89}
would be key for understanding the emission mechanism. Relating
this to the observed coherence of the oscillations, $Q\sim20-50$, could provide an
additional, powerful constraint.

\acknowledgements
I am indebted to Phil Arras and Lars Bildsten for lending their advice
and expertise during many helpful discussions. I also thank Omer Blaes and
Philip Chang for their help.
This work was supported by the National Science Foundation
under grant AST02-05956 and by the Joint Institute for
Nuclear Astrophysics through NSF grant PHY02-16783.


\end{document}